\documentclass[10pt,a4paper,twocolumn,english,prb,aps,showpacs,amsmath,amssymb]{revtex4}
\usepackage{graphicx}
\usepackage[latin9]{inputenc}
\usepackage{url}
\usepackage[update,prepend]{epstopdf}
\usepackage{color}
\usepackage{babel}
\usepackage{amsmath}
\usepackage{amssymb}
\usepackage[unicode=true,
 bookmarks=true,bookmarksnumbered=false,bookmarksopen=false,
 breaklinks=false,pdfborder={0 0 1},backref=false,colorlinks=true]
 {hyperref}
\hypersetup{pdftitle={Pseudogap in high-temperature superconductors from realistic Fr\"ohlich and Coulomb interactions},pdfauthor={A. S. Alexandrov, J. H. Samson, G. Sica}, pdfkeywords={polaron, bipolaron, t-Jp-U model, EPI interaction, NS junction, tunneling, conductance, thermodynamics, specific heat, entropy},linkcolor=blue,urlcolor=blue}

\makeatletter

\usepackage{dcolumn}
\usepackage{bm}

\makeatother

\begin{document}

\title{Pseudogap in high-temperature superconductors from realistic Fr\"ohlich and Coulomb interactions}

\author{G. Sica$^{1,2}$}
\email[Email: ]{gerardo.sica@physics.unisa.it}
\author{J. H. Samson$^{2}$}
\author{A. S. Alexandrov$^{2,3}$}

\address{
$^{1}$ Dipartimento di Fisica ``E.R. Caianiello'', Universit\`{a} degli Studi di Salerno, I-84084 Fisciano (SA), Italy}
\address{$^{2}$ Department of Physics, Loughborough University, Loughborough LE11 3TU, United Kingdom}
\address{$^{3}$Instituto de Fisica ``Gleb Wataghin'', Universidade Estadual de Campinas, UNICAMP 13083-970, Campinas, S\~{a}o Paulo, Brasil}

\date{\today}

\begin{abstract}
It has been recently shown that the competition between unscreened Coulomb and Fr\"{o}hlich electron-phonon interactions can be described in terms of a short-range spin exchange $J_p$ and an effective on-site interaction $\tilde{U}$ in the framework of the polaronic $t$-$J_p$-$\tilde{U}$ model. This model, that provides an explanation for high temperature superconductivity in terms of Bose-Einstein condensation (BEC) of small and light bipolarons, is now studied as a charged Bose-Fermi mixture.
Within this approximation, we show that a gap between bipolaron and unpaired polaron bands results in a strong suppression of low-temperature spin susceptibility, specific heat and tunneling conductance, signaling the presence of normal state pseudogap without any assumptions on preexisting orders or broken symmetries in the normal state of the model.
\end{abstract}


\maketitle

Experimental evidence of finite charge/spin pseudogap (PG), namely a depression of the electronic excitation spectrum at a temperature $T^\ast$ well above the critical temperature $T_c$, have been widely advocated as one of the most significant signatures of hidden orders or broken symmetries in the underdoped regime of high-temperature superconductors.
Although a plethora of different techniques have been used for the investigation of the PG, such as tunneling measurements, angle-resolved photoemission spectroscopy (ARPES), nuclear magnetic relaxation (NMR), Raman and neutron scattering \cite{Timusk1999,Damascelli}, a microscopic theoretical description is still missing \cite{TheoreticalReview}.

Early NMR measurements, in which spin excitations are probed, revealed an anomalous depression in the temperature dependence of the Knight shift for underdoped samples of YBCO \cite{NMRYBCO}. Such a depression, well-known in BCS-like superconductors as induced by the formation of a spin-singlet state, allows some researcher to maintain that PG features could be explained in terms of preformed Cooper pairs\cite{Emery1995,Emery1997} or as signatures of the suppression of low-energy antiferromagnetic fluctuations \cite{AFfluct1,AFfluct2,AFfluct3}.
However, a later interpretation of NMR data as a consequence of a depression in the electron density of states (DOS), results in a number of controversial debates mainly based on the awareness that, in principle, any instability (e.g. stripes, charge/spin density waves, polaron formation) might result in an energy gap. Therefore, the conclusion that pseudogap features might not necessarily imply spin-singlet formation, paves the way to other theories and phenomenological models in which PG emerges as a consequence of SU(2) rotation \cite{SU2}, coexistence of charge and spin density waves\cite{CDWSDW}, inhomogeneous charge distributions \cite{Inhomogeneity1,Inhomogeneity2}.

Recently, strong evidence of particle-hole symmetry breaking in the pseudogap state of Bi2201 shed doubts about the possibility to consider the pseudogap as a precursor of a Cooper pairing superconducting gap in the normal state \cite{PHviolation}. Also some earlier and more recent ARPES experiments \cite{EPIpseudogap} emphasized the role of electron-lattice coupling as an unavoidable ingredient for the characterization of both normal and superconducting state of high-temperature superconductors. Actually, on the phenomenological level, the pseudogap was originally explained as half of the bipolaron binding energy\cite{HolePairs}.

In this context, we report our study on the pseudogap in the framework of the microscopic $t$-$J_p$-$\tilde{U}$ model:
\begin{eqnarray}
\mathcal{H}
&=& -\sum_{i,j}t_{ij}\delta_{\sigma\sigma^{\prime}}c_{i}^{\dagger}c_{j} +\tilde{U} \sum_{\textbf{m}}n_{\textbf{m}\uparrow}n_{\textbf{m}\downarrow}+ \cr
&&+2\sum_{\textbf{m}\neq\textbf{n}}J_{p}(\tilde{U},\textbf{m}-\textbf{n})\left(\textbf{S}_\textbf{m}\cdot\textbf{S}_\textbf{n}+\frac{1}{4}n_{\textbf{m}} n_{\textbf{n}}\right)
\label{tJpU}
\end{eqnarray}
that accounts for realistic Coulomb repulsion and strong electron-phonon (Fr\"{o}hlich) interaction in terms of an exchange coupling $J_p$ \cite{Alexandrov2011,tJpmodel} and a residual on-site correlation $\tilde{U}$ which, limiting the double occupancy, reduces $J_p$ from its bare value \cite{tJpUmodel}. Here $c_i$, $c^\dagger_i$ are polaron annihilation and creation operators where $i=(\textbf{m},\sigma)$ and $j=(\textbf{n},\sigma^{\prime})$ include both site $(\textbf{m},\textbf{n})$ and spin $(\sigma,\sigma^{\prime})$ indices and the sum over $\textbf{n}\neq\textbf{m}$ counts each pair once only. $n_{\textbf{m}}=n_{\textbf{m}\uparrow}+n_{\textbf{m}\downarrow}$, and $n_{\textbf{m}\uparrow, \downarrow}=c^\dagger_{\textbf{m}\uparrow, \downarrow}c_{\textbf{m}\uparrow, \downarrow}$ are site occupation operators and $\textbf{S}_\textbf{m}=(1/2)\sum_{\sigma,\sigma^\prime}c^\dagger_{\textbf{m}\sigma}\overrightarrow{\tau}_{\sigma\sigma^\prime}c_{\textbf{m}\sigma^\prime}$ is the spin $1/2$ operator ($\overrightarrow{\tau}$ are the Pauli matrices). It is worth recalling that our $t$-$J_p$-$\tilde{U}$ model describes carriers doped into the charge-transfer Mott-Hubbard (or any polar) insulator, rather than the insulator itself, different from the conventional Hubbard $U$ or $t$-$J$ models. The bare Hubbard-$U$ on the oxygen orbitals (where doped holes reside) in a rigid cuprate lattice is of the same order of magnitude as the on-site attraction induced by the Fr\"{o}hlich EPI ($\approx 1$eV to $2$eV) so that the residual Hubbard $\tilde{U}$ could be as large as a few hundred meV \cite{tJpUmodel}. Hereafter we restrict our analysis to nearest-neighbor $J$ and $t$ on a square lattice.

We show that, without any ad-hoc assumption on the relative strength and the range of Coulomb and electron-phonon interactions, PG features naturally appear as a consequence of a thermal-induced mixture of polarons and bipolarons, providing a non-Fermi liquid description for the normal state.

\begin{figure}[thb]
\includegraphics[width=1.0\columnwidth]{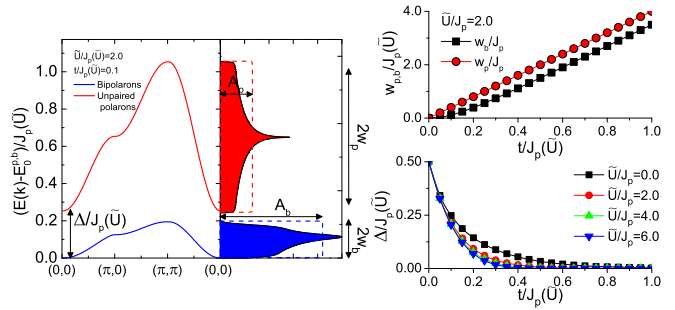}
\caption{(Color online) Left panel: bipolaron (bottom) and unpaired polaron (top) bands with the corresponding DOS: $N_{p,b}(E)$ (filled area) for $\tilde{U}/J_p(\tilde{U})=2.0$ and $t/J_p(\tilde{U})=0.1$. Here $E_0^b=2E_0^p=E_0$, where $E_0$ is the two-particle ground state energy. Dashed lines on the DOS represent the resulting Heaviside-theta approximation. Right panels: DOS parameters versus $t/J_p(\tilde{U})$ for different values of the ratio $\tilde{U}/J_p(\tilde{U})$.}
\label{Fig:DOSapproximation}
\end{figure}

Early numerical and analytical studies on the $t$-$J_p$ and $t$-$J_p$-$\tilde{U}$ models pointed out that the ground state of the system can be described as a coherent superfluid of inter-site small bipolarons\cite{tJpmodel,tJpUmodel}. As follows from a straightforward calculation of the static ($t=0$) ground state configuration, bipolarons repel each other due to the presence of a short-range effective bipolaron-bipolaron repulsion $E_{bb}=(2-\sqrt{3})J_p(\tilde{U})$. Furthermore, they do not attract single polarons.  Hence there is no tendency to clustering. At this point, it is convenient to apply a potential shift to the Hamiltonian: $\mathcal{H} \rightarrow \mathcal{H} - \frac{1}{2}E_0 \sum_\textbf{m} n_\textbf{m}$. Here $E_0$ is the two-particle ground state energy; for small $t$, $E_0 = -J_p$. Because of the aforementioned bipolaron-bipolaron repulsion, all energy eigenstates of the shifted Hamiltonian are non-negative.  The shift has no physical effect, as it will be absorbed into the chemical potential, but provides a more intuitive visualization of the spectrum (Fig.\ref{Fig:DOSapproximation}).

As long as we are in the low-density regime, bipolaron and unpaired polaron interactions are negligible and the model can be described in terms of an ideal (non-interacting) Bose-Fermi mixture of bipolarons (bosons) and unpaired polarons (fermions) in which bipolaron and unpaired polaron densities $n_{p,b}(T)$ are related to the doping $x\ll1$ as\cite{alebook}:
\begin{equation}
  \label{Eq:TotalNumberOfParticle}
  x=2n_b(T,\mu_b)+n_p(T,\mu_p)\;.
\end{equation}
Here $\mu_{p,b}$ represents the polaronic/bipolaronic chemical potential. Charge conservation allows us to fix the chemical potential $\mu$ of the whole mixture. In particular, according to detailed equilibrium and total energy conservation one readily obtains $\mu=\mu_p=\mu_b/2$. Then the thermodynamic properties of the Fermi-Bose mixture at equilibrium can be easily derived in terms of the thermodynamic potential $\Omega(x,T)$, defined as a function of temperature $T$ and chemical potential $\mu$ \cite{note}.

In the dilute limit all the thermodynamic quantities enjoy the additivity property therefore $\Omega(x,T)$ can be expressed as the sum of bipolaron/unpaired polarons potentials $\Omega_{b,p}(n_{b,p},T)$ with:
\begin{equation}\small
\label{Eq:Omega}
  \Omega_{p,b}=\mp k_{B}T\int_{_{-\infty}}^{^{+\infty}}d\epsilon\mathcal{N}_{p,b}(\epsilon)\ln\left(1\pm\exp\left(\frac{\mu_{p,b}-\epsilon}{k_{B}T}\right)\right)\;.
\end{equation}
Here $\mathcal{N}_{b,p}(\epsilon)$ is the density of states in the bipolaron/unpaired polaron band, respectively. For the sake of simplicity, hereafter the DOS of the model is approximated as:
\begin{equation}
  \label{Eq:DOSapproximation}
  \begin{cases}
    \mathcal{N}_{p}(\epsilon)=A_{p}\Theta\left(\epsilon-\Delta\right)\Theta\left(\Delta+2w_{p}-\epsilon\right)\\
    \mathcal{N}_{b}(\epsilon)=A_{b}\Theta(\epsilon)\Theta(2w_{b}-\epsilon)
  \end{cases}\;.
\end{equation}
In a square lattice with nearest-neighbour $J_p$, there are two bound bosonic states per site, corresponding to the two bonds per site and two fermionic states per site due to spin degeneracy. Thus rectangular bands (including the degeneracy factors) need to satisfy the constraint $\int \mathcal{N}_{p,b}(\epsilon)d\epsilon=2$ from which we have $A_{p,b}=1/w_{p,b}$. Here $A_{p,b}$ and $w_{p,b}$ are intensity and half-bandwidth of polaron and bipolaron terms, respectively, $\Delta$ represents the binding energy per polaron.

Importantly, DOS and model parameters are linked together: $\Delta\approx J_p(\tilde{U})/2$ at $t/J_p\ll1$ while bandwidths $w_{p,b}$ are related to the ratio $t/J_p(\tilde{U})$ as reported in Fig.\ref{Fig:DOSapproximation}. Such an approximation is able to get an insight in a qualitative microscopic description of the pseudogap. It allows to obtain analytical expressions for the relevant physical properties. In particular, bipolaron and polaron densities, calculated as $\int_{-\infty}^{\infty}d\epsilon\mathcal{N}_{p,b}(\epsilon)f_{p,b}(\epsilon,T)$, where $f_{p,b}(\epsilon,T)=[\exp{(\epsilon-\mu_{p,b})/k_BT}\pm1]^{-1}$ is the Fermi-Dirac/Bose-Einstein distribution function, are expressed as:
\begin{equation}
\label{Eq:npb}
\small
  \begin{split}
    n_{b}(T)=&-1+\frac{k_BT}{w_b}\ln\left[\frac{\sinh\left(\frac{w_{b}-\mu}{k_{B}T}\right)}{\sinh\left(-\frac{\mu}{k_{B}T}\right)}\right]\;,\\
    n_{p}(T)=&1-\frac{k_BT}{w_p}\ln\left[\frac{\cosh\left(\frac{\Delta+2w_{p}-\mu}{2k_{B}T}\right)}{\cosh\left(\frac{\Delta-\mu}{2k_{B}T}\right)}\right]\;,
  \end{split}
\end{equation}
from which $\mu$ can be calculated self-consistently according to Eq.\ref{Eq:TotalNumberOfParticle}.

\begin{figure}[tbp]
\begin{tabular}{cc}
\includegraphics[width=0.5\columnwidth]{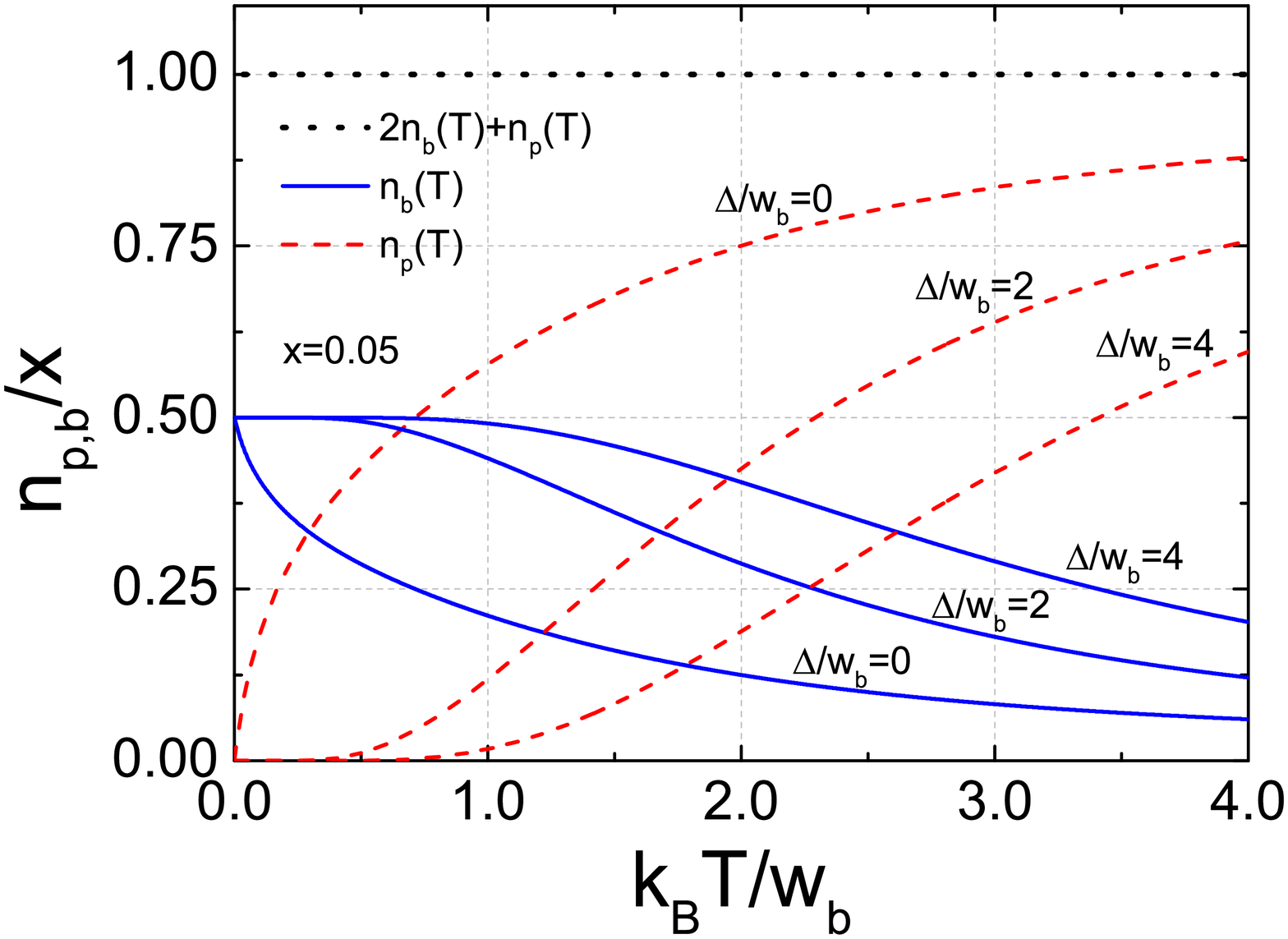}
\includegraphics[width=0.5\columnwidth]{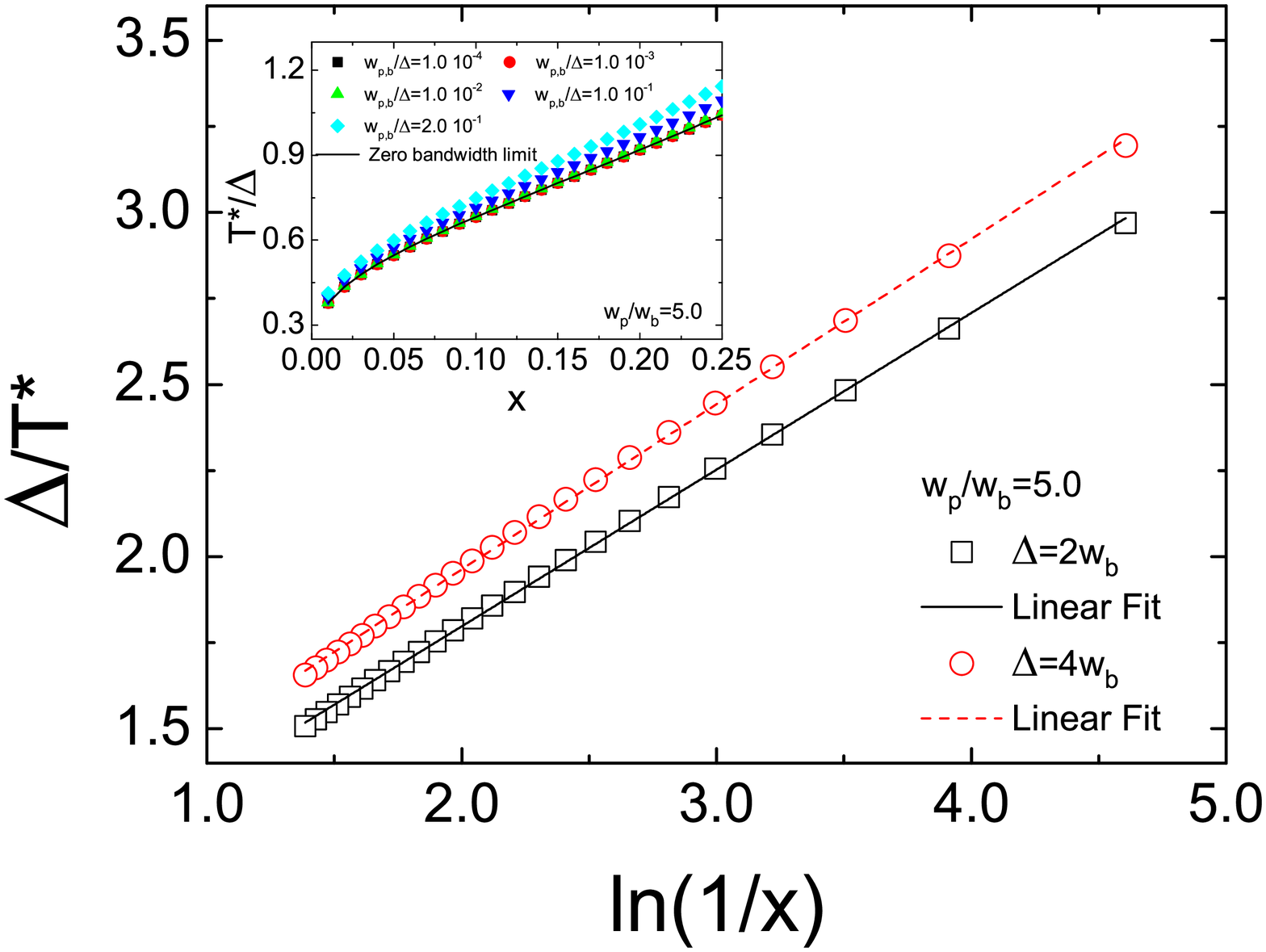}
\end{tabular}
\caption{(Color online) Left panel: relative bipolaron (solid)/unpaired polaron (dashed) density versus temperature for different values of the gap $\Delta$. The dotted line represents the total particle density $x=2n_b+n_p$. Right panel: linear dependence of the ratio $\Delta/T^\ast$ with respect to $\ln(1/x)$ for different value of the gap $\Delta$. In the inset the doping dependence of $T^\ast/\Delta$ (symbols) is compared with the exact analytical dependence (line) obtained in the zero-bandwidth limit. Here $T^\ast$ is the crossover temperature at which $n_b(T^\ast)=2n_p(T^\ast)$.}
\label{Fig:MuNpb}
\end{figure}

Our results on chemical potential and particle density, reported in Fig.\ref{Fig:MuNpb} for a fixed total number of particle $x$, show that different temperature behaviors arise in polaron and bipolaron densities depending on the value of $\Delta$ and on the competition between the pairing interaction $J_p(\tilde{U})$ and hopping term (Fig.\ref{Fig:DOSapproximation}). In particular, as follows from the left panel of Fig.\ref{Fig:MuNpb}, the bipolaron density decreases with increasing temperature resulting in a crossover at $T=T^\ast$ when half of the bipolarons are dissociated and the charge is equally distributed between polarons and bipolaron ($n_b(T^\ast)=2n_p(T^\ast)$). As shown in the right panel of Fig.\ref{Fig:MuNpb} the ratio $\Delta/k_BT^\ast$ varies linearly with $\ln(1/x)$ in a wide range of doping, with $k_BT^\ast=2\Delta/\ln((4/x-1)^2/(1+8/x))$ for $w_{p,b}/\Delta\rightarrow0$ in agreement with exact analytical calculations in the narrow-band limit.

The thermal-induced population of the unpaired polaron band results in a number of anomalous features that can be observed in specific heat $C(T)$ and static uniform spin susceptibility $\chi_s(T,h)$ ($h$ is an external magnetic field). Recalling that $C(T)=d\langle E\rangle/dT$, where $\langle E\rangle=\langle E_p\rangle+\langle E_b\rangle$ is the total energy of the system with $\langle E_{p,b}\rangle=\int d\epsilon E\mathcal{N}_{p,b}(\epsilon)f_{p,b}(\epsilon,T)$, and $\chi_s(T,h)=-\partial^2\Omega/\partial h^2$, we have:
\begin{equation}\small
  \label{Eq:ChiS}
  \chi_{s}(h,T)=\frac{\mu_{B}^{2}}{2k_{b}T}\int_{-\infty}^{+\infty}d\epsilon\mathcal{N}_{p}(\epsilon)\frac{1+\cosh\left[\frac{\epsilon-\mu_{p}}{k_{B}T}\right]\cosh\left[\frac{\mu_{B}h}{k_{B}T}\right]}{\left(\cosh\left[\frac{\epsilon-\mu_{p}}{k_{B}T}\right]+\cosh\left[\frac{\mu_{B}h}{k_{B}T}\right]\right)^{2}}\;,
\end{equation}
\begin{equation}\small
  \label{Eq:Cv}
  C(T)=k_BT\sum_{k={b,p}}A_k\left[k_BI_k^{(2)}(x)+\frac{d\mu_n}{dT}I^{(1)}_k(x)\right]_{x_k^{in}}^{x_k^{fin}}\;.
\end{equation}
Here $\mu_B$ is the Bohr magneton and:
\begin{equation}\small
  \begin{cases}
    x_{b}^{in}=-\frac{\mu_{b}}{k_{B}T}\\
    x_{b}^{fin}=\frac{2w_{b}-\mu_{b}}{k_{B}T}
  \end{cases}\;,\;
  \begin{cases}
    x_{p}^{in}=\frac{\Delta-\mu_{p}}{k_{B}T}\\
    x_{p}^{fin}=\frac{\Delta+2w_{p}-\mu_{p}}{k_{B}T}
  \end{cases}\;,
\end{equation}
while $I^{(n)}_{p,b}(x)\equiv\int dx\frac{x^{n}e^{x}}{\left(e^{x}\pm1\right)^{2}}$ is expressed in terms of the polylogarithm function $Li_s(z)=\sum_{k=1}^{\infty}z^k/k^s$ as:
\begin{equation}\small
  \label{Eq:In}
  I_{p,b}^{(n)}(x)=
  \begin{cases}
    \frac{xe^{x}}{1\pm e^{x}}\mp\ln\left(1\pm e^{x}\right) & \;,\; n=1\\
    x\left(\frac{xe^{x}}{1\pm e^{x}}\mp2\ln\left(1\pm e^{x}\right)\right)\mp Li_{2}\left(\mp e^{x}\right) & \;,\; n=2
  \end{cases}\;.
\end{equation}

\begin{figure}[tbp]
\begin{tabular}{cc}
\includegraphics[width=0.5\columnwidth]{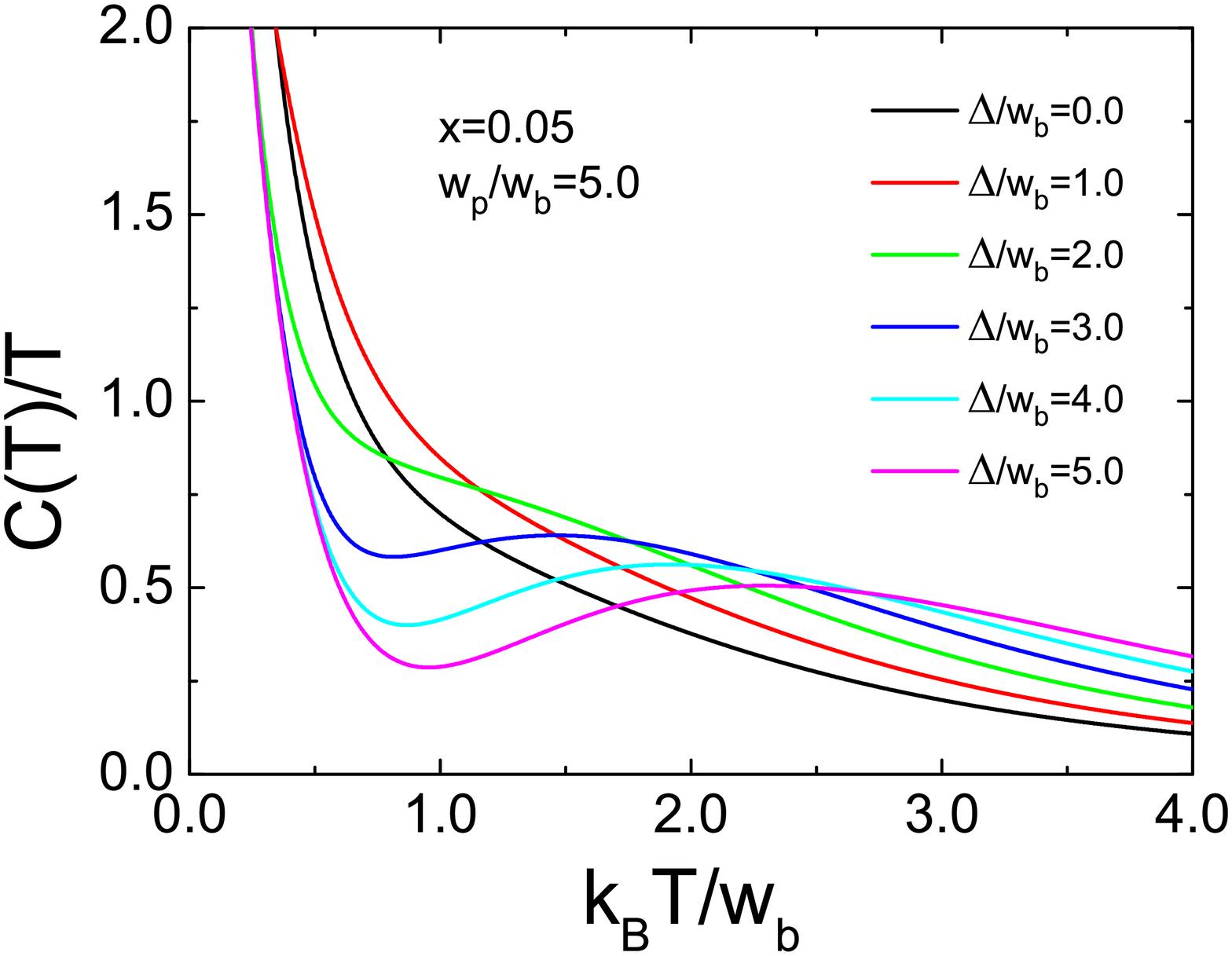}
\includegraphics[width=0.5\columnwidth]{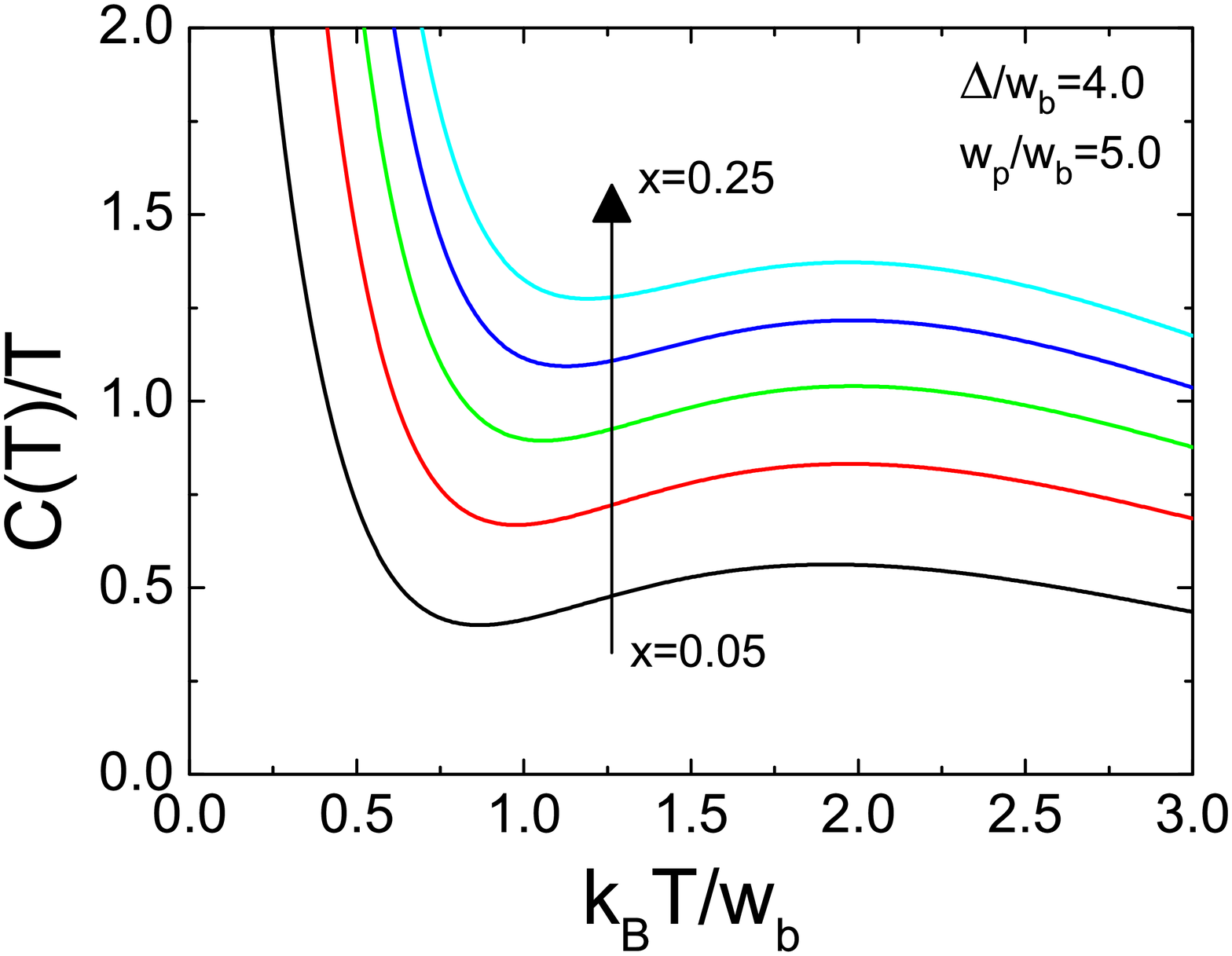}
\end{tabular}
\caption{(Color online) Specific heat coefficient $\gamma(T)=C(T)/T$ versus temperature plotted for different values of gap $\Delta$ (left panel) and doping $x$ (right panel). Here $w_{b,p}$ is the half-bandwidth of the bipolaron/unpaired polaron band.}
\label{Fig:Cv}
\end{figure}

\begin{figure}[tbp]
\begin{tabular}{cc}
\includegraphics[width=1.0\columnwidth]{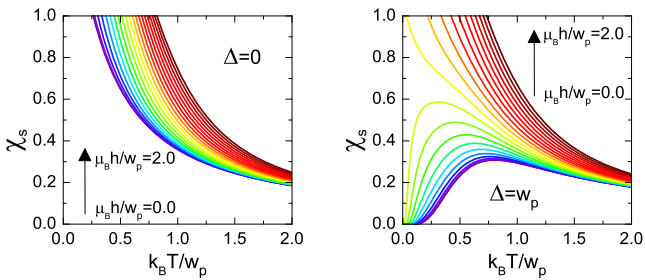}
\end{tabular}
\caption{(Color online) Spin susceptibility (Eq.\ref{Eq:ChiS}) versus temperature plotted for different values of the external magnetic field $h$. Here $\Delta$ is the gap between bipolaron and unpaired polaron bands, $w_p$ is the half-bandwidth of the polaronic band.}
\label{Fig:ChiS}
\end{figure}

As follows from Fig.\ref{Fig:DOSapproximation}, for $t/J_p(\tilde{U})\approx1$ and for $\tilde{U}\gg1$ ($J_p(\tilde{U}\rightarrow\infty)\rightarrow0$\cite{tJpUmodel}), the gap $\Delta$ goes to zero. In this case bipolaron and unpaired polaron bands are completely overlapped with a single peak in the specific heat coefficient $\gamma(T)=C(T)/T$ leading to $\gamma(T)\propto1/T$, Fig.\ref{Fig:Cv}. Consistently, the paramagnetic response of the system reproduces the standard Curie law with $\chi_{s}(T,h=0)\propto1/T$ (Fig.\ref{Fig:ChiS}) since fermions are non-degenerate under the conditions here.

On the contrary, in the opposite regime the presence of a finite gap results in a non monotonic $\gamma(T)$ dependence induced by the superposition of two main peaks due to intra-band and bipolaron to unpaired polarons excitations. While any finite temperature can induce intra-band excitations, bipolaron dissociation requires temperatures of the order of the gap. Therefore the separation between the two peaks increases with increasing $\Delta$ and results in a strong suppression of the $\gamma(T)$ coefficient in the region in which the intensity of the intra-band peak falls off ($k_BT\approx w_b$).
Specific heat and static uniform spin susceptibility are both determined by the average density of electronic states therefore, as one would expect, the same tendency is also observed in the paramagnetic response function $\chi_s(T,h=0)$. In fact, as confirmed in Fig.\ref{Fig:ChiS}, $\chi(T,h=0)$ drops to zero in the low temperature regime in which the population of the bipolaronic (singlet) band has its maximum and follows the standard Curie law, already described in the $\Delta=0$ case, in the high-temperature regime in which the population of the polaronic band become dominant. The same features also appears in the presence of an external magnetic field until $\mu_Bh\approx\Delta$. For $\mu_Bh>\Delta$ the magnetic field induces a finite magnetization in the system with a singlet/triplet phase transition at $\mu_Bh=\Delta$ signalled by a discontinuity in the spin susceptibility at $T=0$.

Finally, let us discuss the tunneling conductance of a normal metal-bipolaronic superconductor (NS) junction:
\begin{equation}\small
  \label{Sigma}
  \sigma_{NS}(V)=\frac{dI_{NS}(V)}{dV}\;,
\end{equation}
in which $V$ is the bias and $I_{NS}(V)$ is the current flowing through the junction. According to the theory of extrinsic and intrinsic tunneling in bosonic superconductors\cite{Conductance1,Conductance2}, $I_{NS}(V)$ can be calculated starting from the following tunneling Hamiltonian:
\begin{equation}\small
  H_{_{NS}}=P\sum_{\nu\nu'}\left(p_{\nu'}^{\dagger}c_{\nu}+c^\dagger_{\nu}p_{\nu'}\right)+\frac{B}{\sqrt{N}}\sum_{\nu\nu'\eta'}\left(b_{\eta'}^{\dagger}p_{\nu'}c_{\nu}+c^\dagger_{\nu}p^\dagger_{\nu'}b_{\eta'}\right)\;.
\end{equation}
that accounts for single particle tunneling processes. Here $c_\nu$, $p^\dagger_{\nu^\prime}$ and $b_{\eta}^{\dagger}$ describe the annihilation of a carrier in the metallic tip in state $\nu$ and the creation of a single polaron or a composed boson in the superconductor in state $\nu^\prime$ or $\eta$ respectively, $N$ is the number of lattice cells. $P$ and $B$ are tunneling matrix elements respectively with and without the involvement of a bipolaron (generally $B\geq P$  \cite{Conductance1}). The tunneling current is $I_{NS}(V)=e\left(W_{N\rightarrow S}-W_{S\rightarrow N}\right)$ where $W_{X\rightarrow Y}$ represents the tunneling probability of transition, per unit time, from the $X$ to the $Y$ side of the junction. According to the Fermi golden rule we have:
\begin{equation}
\small
\begin{split}
I_{NS}(V)=&\frac{2\pi e}{\hbar}A_{m}A_{p}
\Biggl\{|P|^2\int_{\Delta}^{\Delta+2w_{p}}d\xi^{\prime}\left[f_F(\xi^{\prime}-eV)-f_{F}(\xi^{\prime})\right]\\
&+|B|^2A_{b}\int_{0}^{2w_{b}}d\eta\int_{\Delta}^{\Delta+2w_{p}}d\xi^{\prime}f_F(\eta-\xi'-eV)f_{F}(\xi^{\prime})\\
&-|B|^2A_{b}\int_{0}^{2w_{b}}d\eta\int_{\Delta}^{\Delta+2w_{p}}d\xi'f_{B}(\eta)\Bigl(1-f_{F}(\xi')\\
&\qquad\qquad-f_F(\eta-\xi'-eV)\Bigr)\Biggr\}\;,
\end{split}
\label{Eq:TunnelingCurrent}
\end{equation}
where $f_{F}(\xi)=\left[\exp\left(\xi/k_{B}T\right)+1\right]^{-1}$ is the Fermi distribution functions associated to normal metal and polaronic band; $f_{B}(\eta)=\left[\exp\left(\eta/k_{B}T\right)-1\right]^{-1}$ represents the Bose distribution function associated to the bipolaronic band. Here we have used a constant DOS for the normal metal with $\mathcal{N}_m(\epsilon)=A_m$.

\begin{figure}[tbp]
\begin{tabular}{cc}
\includegraphics[width=0.50\columnwidth]{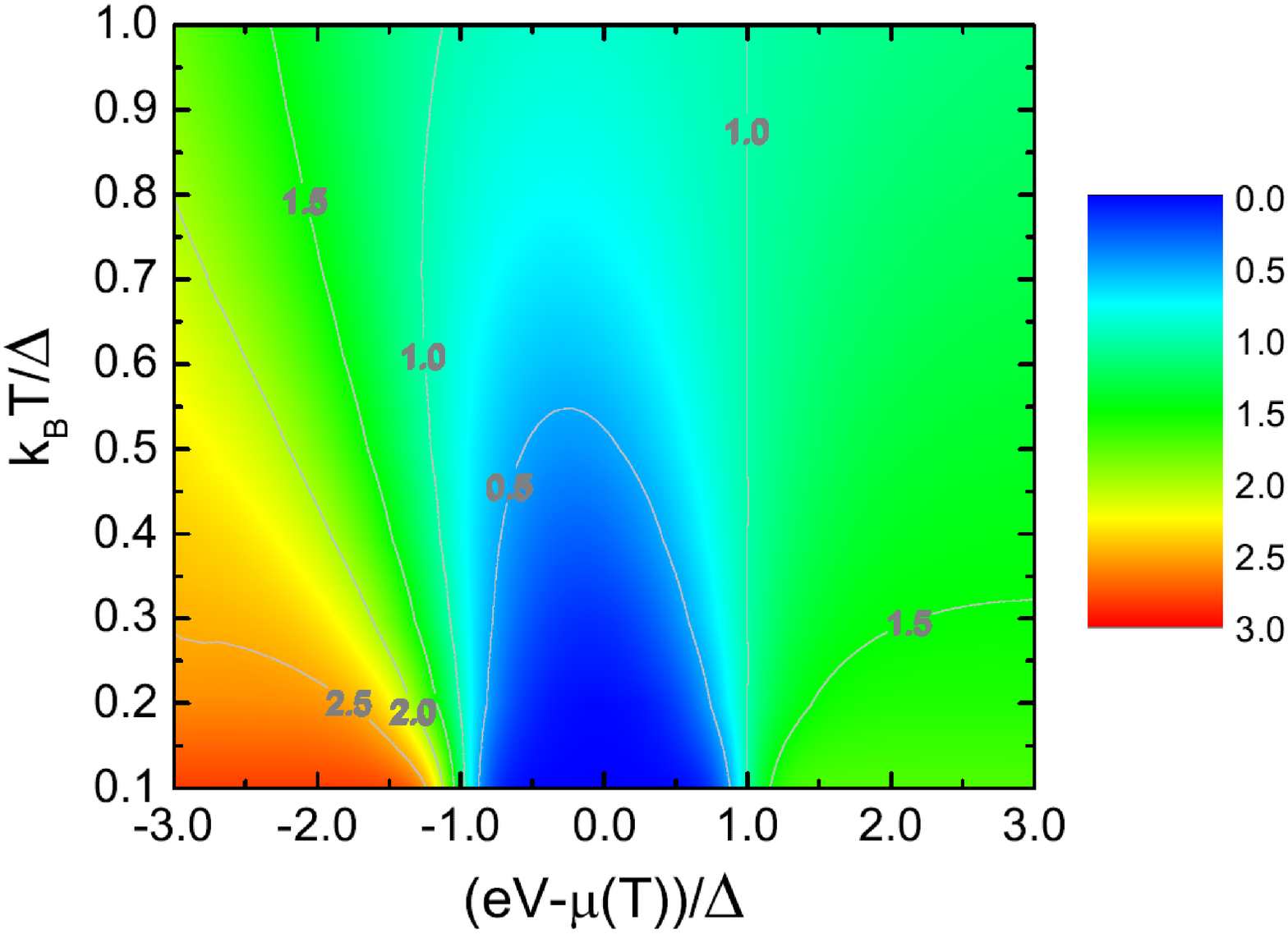}
\includegraphics[width=0.40\columnwidth]{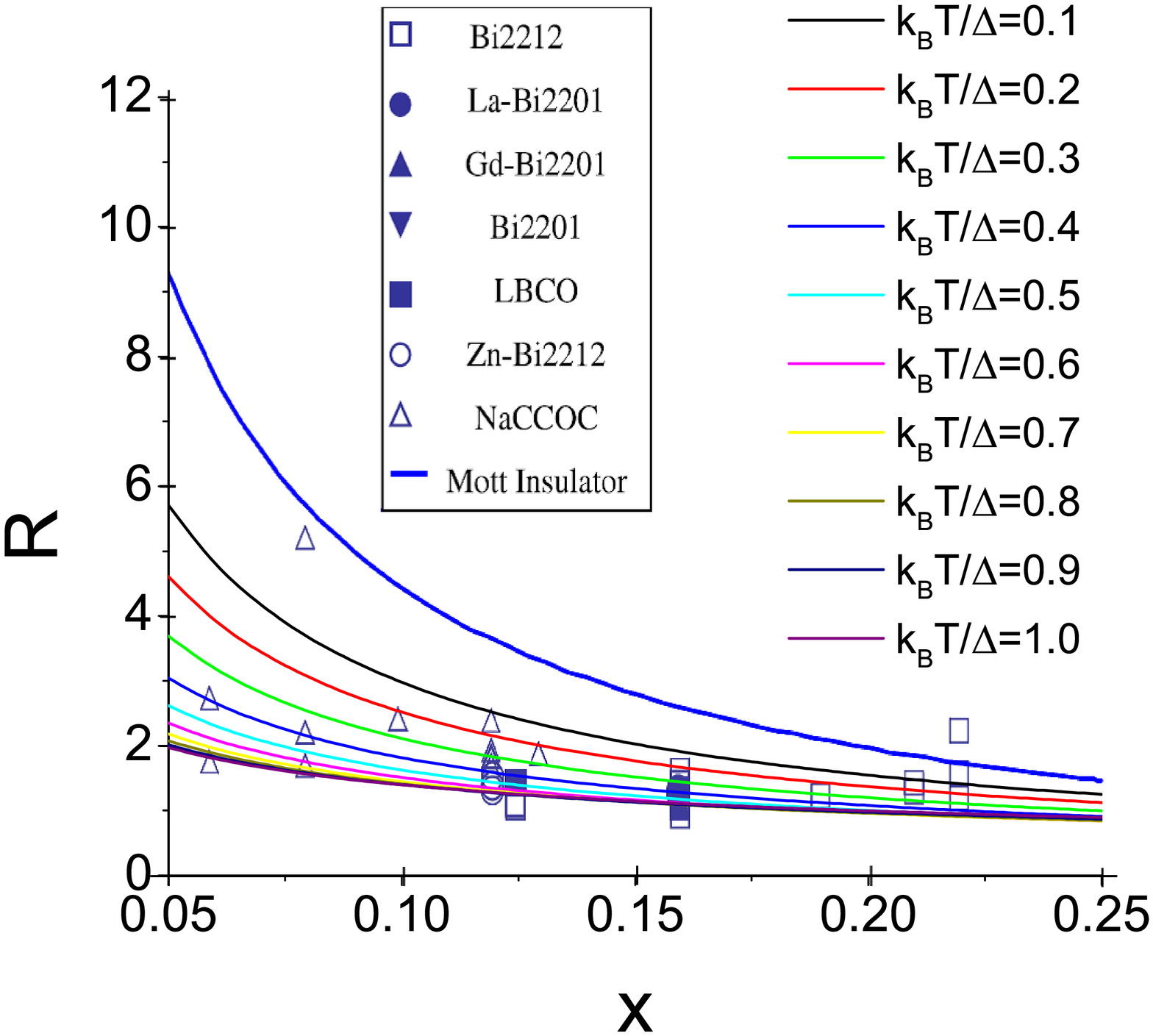}
\end{tabular}
\caption{(Color online) Left panel: density plot of the normalized conductivity $\sigma_{NM}(V)/\sigma_{NM}(\Delta)$ in the $k_BT/\Delta$ - $(eV-\mu(T))/\Delta$ plane. Right panel: doping dependence of the asymmetry coefficient $R(x,T)$, Eq.\ref{Eq:Rcoefficient}. Numerical results obtained by integrating the normalized conductivity $\sigma(eV)/\sigma(\Delta)$ from $0$ to $\pm\Delta$ for different values of the temperature are compared with experimental results in cuprates (from Ref.\onlinecite{Conductance2}).}
\label{Fig:Conductance}
\end{figure}

As reported in Fig.\ref{Fig:Conductance}, in the same regime in which pseudogap features arise in the specific heat (Fig.\ref{Fig:Cv}) and spin susceptibility (Fig.\ref{Fig:ChiS}), our results on the tunneling conductivity confirm a strong depression of $\sigma(V)$ at zero bias for $k_{B}T\leq\Delta$. Importantly, despite the lack of van Hove singularities in the DOS (Fig.\ref{Fig:DOSapproximation}), our data perfectly reproduce the asymmetry between negative and positive bias conductance providing a further confirmation that the van Hove singularity, not observed in many experiments such as momentum integrated photoemission \cite{vanHove}, does not play any role in the tunneling.
Our model allows us to describe the doping dependence of the asymmetry coefficient $R(x,T)$ defined as:
\begin{equation}
\label{Eq:Rcoefficient}
  \small
  R(x,T)=\frac{\int_{-\Delta}^{0}\sigma(eV)dV}{\int_{0}^{\Delta}\sigma(eV)dV}=\frac{I_{NS}(-\Delta)}{I_{NS}(\Delta)}\;.
\end{equation}
We recall that with increasing doping the bipolaron density increases while the unpaired polaron density remains almost zero for $k_BT<w_b$ (see Fig.\ref{Fig:MuNpb}, left panel). In this regime the polaronic contribution to the tunneling conductance remains constant while the bipolaron contribution, different from zero only in the positive bias regime, scales linearly with the bipolaron density as clearly follows from Eq.\ref{Eq:TunnelingCurrent} if one neglects the bipolaron energy dispersion in the narrow bipolaron-band limit. As reported in Fig.\ref{Fig:Conductance}, this explanation is also supported by a good agreement between our data and experimental measurements of the asymmetry coefficient $R(x,T)$ in a wide range of cuprates superconductors.
Importantly, it is worth noting that numerical data for $R(x,T)$ have been calculated by integrating the normalized conductivity $\sigma(eV)/\sigma(\Delta)$, therefore do not depend on the particular choice of the tunneling matrix elements $B$, and $P$ in Eq.\ref{Eq:TunnelingCurrent}. As one would expect, the only relevant quantities are $t$, $J_p(\tilde{U})$, $\tilde{U}$ that, unlike in other theories, in the $t$-$J_p$-$\tilde{U}$ model can be fixed from the measurable material properties\cite{Alexandrov2011,tJpmodel,tJpUmodel}.

In conclusion, we have described the normal state of the polaronic $t$-$J_p$-$\tilde{U}$ model as an ideal Bose-Fermi mixture in the low density limit. By approximating the DOS of the model as rectangular functions \eqref{Eq:DOSapproximation}, we have provided analytical expressions for bipolaron and unpaired polaron densities \eqref{Eq:npb} and most of the relevant response functions such as specific heat \eqref{Eq:Cv}, spin susceptibility \eqref{Eq:ChiS}. Our analysis pointed out that in the presence of a finite gap $\Delta$ between bipolaron and unpaired polaron bands, the model exhibits remarkable features of pseudogap opening signaled by a depression of specific heat, spin susceptibility and tunneling conductance. Furthermore, as the result of the screening\cite{AleBratk2010,AleKabMott1996}, the pseudogap $\Delta$ falls with doping through the dielectric function so that the crossover temperature of our model falls as well. Importantly, different  from any other theories proposed so far, pseudogap features naturally appear as a consequence of a thermal-induced mixture of polarons and bipolarons, without any ad-hoc assumption on relative strength and the range of Coulomb and electron-phonon interactions or preexisting orders.

We gratefully acknowledge Adolfo Avella, Roberta Citro, Mario Cuoco, Jim Hague, Ferdinando Mancini, Evgeny Plekhanov, Maurice Rice for stimulating discussions and the UNICAMP vi\-si\-ting professorship program (Brasil).



\begin{thebibliography}{99}
\bibitem{Timusk1999} For a review see T. Timusk and B. Statt, \href{http://link.aps.org/doi/10.1103/PhysRevLett.106.136403}{Rep. Prog. Phys. \textbf{62}, 61 (1999)}.
\bibitem{Damascelli} For a review see A. Damascelli, Z. Hussain, Z.-X. Shen, \href{http://link.aps.org/doi/10.1103/RevModPhys.75.473}{Rev. Mod. Phys. \textbf{75}, 473 (2003).}
\bibitem{TheoreticalReview} For a review see M. R. Norman, D. Pines, and C. Kallin, \href{http://dx.doi.org/10.1080/00018730500459906}{Adv. Phys. \textbf{54}, 715 (2005)}.
\bibitem{NMRYBCO}W. W. Warren, R. E. Walstedt, G. F. Brennert, R. J. Cava, R. Tycko, R. F. Bell, and G. Dabbagh, \href{http://link.aps.org/doi/10.1103/PhysRevLett.62.1193}{Phys. Rev. Lett. \textbf{62}, 1193 (1989)}.

\bibitem{Emery1995} V. J. Emery, and S. A. Kivelson, \href{http://dx.doi.org/10.1038/374434a0}{Nature \textbf{374}, 434 (1995)}.
\bibitem{Emery1997} V. J. Emery, S. A. Kivelson, and O. Zachar, \href{http://link.aps.org/doi/10.1103/PhysRevB.56.6120}{Phys. Rev. B \textbf{56}, 6120 (1997)}.

\bibitem{AFfluct1} D. C. Johnston, \href{http://link.aps.org/doi/10.1103/PhysRevLett.62.957}{Phys. Rev. Lett. \textbf{62} 957 (1989)}.
\bibitem{AFfluct2} T. Nakano, M. Oda, C. Manabe, N. Momono, Y. Miura and M. Ido, \href{http://link.aps.org/doi/10.1103/PhysRevB.49.16000}{Phys. Rev. B \textbf{49}, 16000 (1994)}.
\bibitem{AFfluct3} N. Curro, Z. Fisk and D. Pines, \href{http://dx.doi.org/10.1557/mrs2005.121}{MRS Bulletin \textbf{30}(6) 442 (2005)}.

\bibitem{SU2} X. G. Wen, and P. A. Lee, \href{http://dx.doi.org/10.1103/PhysRevLett.76.503}{Phys. Rev. Lett. \textbf{76}, 503 (1996)}.
\bibitem{CDWSDW} B. Pradhan, B. K. Raj, and G. C. Rout, \href{http://dx.doi.org/10.1016/j.physc.2008.08.010}{Physica C \textbf{468}, 2332 (2008)}.

\bibitem{Inhomogeneity1} E. V. L. de Mello, M. T. D. Orlando, J. L. Gonzalez, E. S. Caixeiro, and E. Baggio-Saitovich, \href{http://dx.doi.org/10.1103/PhysRevB.66.092504}{Phys. Rev. B \textbf{66}, 092504 (2002)}.
\bibitem{Inhomogeneity2} V. J. Emery, S. A. Kivelson, and H. Q. Lin, \href{http://dx.doi.org/10.1103/PhysRevLett.64.475}{Phys. Rev. Lett. \textbf{64}, 475 (1990)}.

\bibitem{PHviolation} M. Hashimoto, H.R. He, K. Tanaka, J.P. Testaud, \emph{et al.}, \href{http://dx.doi.org/10.1038/nphys1632}{Nat. Phys. \textbf{6}, 414 (2010)}.

\bibitem{EPIpseudogap} K.M. Shen, F. Ronning, D.H. Lu, W.S. Lee, N.J.C. Ingle, W. Meevasana, F. Baumberger, A. Damascelli, N.P. Armitage, L.L. Miller, Y. Kohsaka, M. Azuma, M. Takano, H. Takagi and Z.-X. Shen, \href{http://link.aps.org/doi/10.1103/PhysRevLett.93.267002}{Phys. Rev. Lett. \textbf{93}, 267002 (2004)}.

\bibitem{HolePairs} A. S. Alexandrov, and D. K. Ray, \href{http://dx.doi.org/10.1080/09500839108214658}{Phil. Mag. Lett. 63, 295 (1991)}.

\bibitem{Alexandrov2011} A. S. Alexandrov, \href{http://dx.doi.org/10.1209/0295-5075/95/27004}{EPL \textbf{95}, 27004 (2011)}.
\bibitem{tJpmodel} A. S. Alexandrov, J. H. Samson, and G. Sica, \href{http://link.aps.org/doi/10.1103/PhysRevB.85.104520}{Phys. Rev. B \textbf{85}, 104520 (2012)}.
\bibitem{tJpUmodel} A. S. Alexandrov, J. H. Samson, and G. Sica, \href{http://arxiv.org/abs/1205.3436v2}{arXiv:1205.3436v2}.

\bibitem{alebook}For a review see A. S. Alexandrov, \textit{Theory of Superconductivity: From Weak to Strong Coupling} (IoP Publishing, Bristol 2003).

\bibitem{note}Apart from the low-density limit, $x\ll1$, we assume that the temperature is above the Bose-Einstein (3D) or BKT (2D) superconducting critical temperature.

\bibitem{AleBratk2010} A. S. Alexandrov and A. M. Bratkovsky, \href{http://link.aps.org/doi/10.1103/PhysRevLett.105.226408}{Phys. Rev. Lett. \textbf{105}, 226408 (2010)}.

\bibitem{AleKabMott1996} A. S. Alexandrov, V. V. Kabanov, and N. F. Mott, \href{http://link.aps.org/doi/10.1103/PhysRevLett.77.4796}{Phys. Rev. Lett. \textbf{77}, 4796 (1996)}.

\bibitem{Conductance1} A. S. Alexandrov and J. Beanland, \href{http://dx.doi.org/10.1103/PhysRevLett.104.026401}{Phys. Rev. Lett \textbf{104}, 026401 (2010)}.
\bibitem{Conductance2} A. S. Alexandrov and J. Beanland, \href{http://dx.doi.org/10.1088/0953-8984/22/40/403202}{J. Phys.: Condens. Matter \textbf{22}, 403202(2010)}.

\bibitem{vanHove} R-H. He, X. J. Zhou, M. Hashimoto \emph{et al.} \href{http://dx.doi.org/10.1088/1367-2630/13/1/013031}{New J. Phys. \textbf{13}, 013031 (2011)}.

\end{thebibliography}
\end{document}